\documentclass{elsart} 

\usepackage{graphicx} 
\usepackage{color}
\usepackage{epsfig}
\usepackage{epstopdf}
  
\usepackage{amssymb}

\begin{document}

\begin{frontmatter}



\title{Cosmogenic Production as a Background in Searching for Rare Physics Processes}

\author[usd]{D.-M. Mei}
\ead{Dongming.Mei@usd.edu},
\author[usd,ccnu,keylab]{Z.-B. Yin\thanksref{now}},
\thanks[now]{Permanent Address: Institute of Particle Physics, 
Huazhong Normal University, Wuhan 430079, China}
\author[lanl]{S. R. Elliott}
%\author[usd]{Chao Zhang}
\address[usd]{Department of Physics, The University of South Dakota, 
Vermillion, South Dakota 57069}
\address[ccnu]{Institute of Particle Physics, Huazhong Normal University, Wuhan 430079, China}
\address[keylab]{Key Laboratory of Quark \& Lepton Physics (Huazhong Normal University), Ministry of Education, China}
\address[lanl]{Los Alamos National Laboratory, Los Alamos, New Mexico 87545}
\begin{abstract}
% Text of abstract
We revisit calculations of the cosmogenic production rates for several long-lived isotopes that are potential sources of background 
in searching for rare physics processes such as the detection of dark matter and neutrinoless double-beta decay. Using updated cosmic-ray neutron flux measurements,
we use TALYS 1.0 to investigate the cosmogenic activation of stable isotopes of
several detector targets and find that the cosmogenic
isotopes produced inside the target materials and cryostat can result in large
backgrounds for dark matter searches and neutrinoless double-beta decay. 
We use previously published low-background HPGe data to constrain the production of  $^{3}H$ on the
surface and the upper limit is consistent with our calculation.
We note 
that cosmogenic production of several isotopes in various targets 
can generate potential backgrounds for dark matter detection and neutrinoless 
double-beta decay 
with a massive detector, thus great care should be 
taken to limit and/or deal with the cosmogenic
activation of the targets.
\end{abstract}

\end{frontmatter}

% main text
\section{Introduction}
\label{sec:intr}

The direct detection of weakly interacting massive particles 
(WIMP)~\cite{Gait04} 
and the search for neutrinoless double-beta ($0\nu\beta\beta$) decay~\cite{Avig07}
are fundamentally important for physics beyond the standard 
model of particle physics. The direct detection of WIMPs 
would help to determine the mass
and cross section of a WIMP, while any observation of neutrinoless double-beta
decay would clearly show that the neutrino is a Majorana particle and that the lepton 
number is not conserved. Both types of 
experiments are searches for extremely rare signals and 
thus require large mass exposures of sensitive detectors with 
small internal backgrounds and sufficient 
shielding against external
backgrounds at a deep underground site. 
%There are proposals to use one 
%common detector to search for both WIMPs and neutrinoless double-beta 
%decay events. Many sources or targets are under consideration that contain isotopes which can undergo 
%double-beta decay.

The physics goals of upcoming double-beta decay experiments are to probe the
quasi-degenerate neutrino mass region as low as 100 meV and demonstrate
that backgrounds can be achieved at or below 1 count/ton/year in the 
$0\nu\beta\beta$ decay peak region of interest (ROI). To realize
these goals, these experiments must construct detectors in an 
ultra-low background structure~\cite{Majo06}. Although these experiments are foremost
neutrino mass experiments, they may potentially  contribute to 
dark matter searches~\cite{MaDa01,Barb07}. In contrast, dark matter experiments
will not necessarily be built of a material composed of $0\nu\beta\beta$ isotope.  In any case, cosmic-ray produced long-lived
isotopes are potential sources of background for either type of experiment. 

 In this paper, we revisit old calculations of the production rates for critical cosmic-ray produced isotopes. The need for this revisit arises, because of recent improvements in the understanding of the cosmic-ray neutron flux and, in the case of $^3$H, the availability of reaction codes that fully identify all the reaction products in the final state. In Section~\ref{sec:bkg} we evaluate,  several long-lived isotopes that are produced in
germanium by cosmic ray neutrons while the material resides on the Earth's surface.  We then validate our calculation
by comparing them to the measured production rates
in Section~\ref{sec:sens}. We discuss the effect of 
cosmogenic activation of natural xenon and various other targets on
dark matter detection in Section~\ref{sec:xe}. 
Finally, we give our conclusions in Section~\ref{sec:con}.

\section{Cosmogenic production in germanium }
\label{sec:bkg}

\subsection{Cosmogenic production of radioactive isotopes}
\label{subsec:cosmo}
\begin{figure}[htb!!!]
\includegraphics[angle=0,width=10.cm] {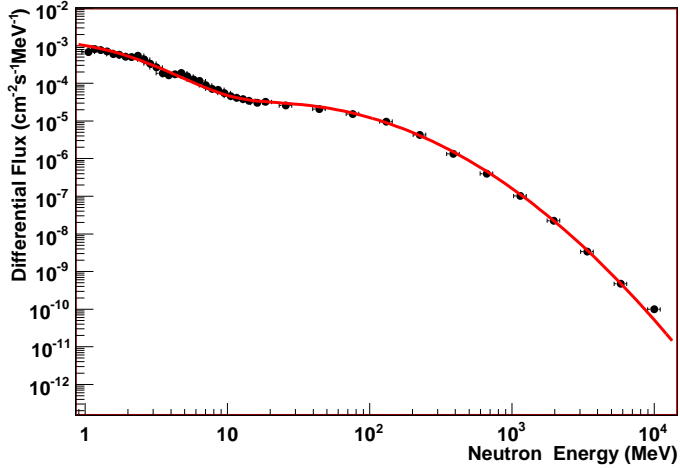}
\caption{\small{The measured neutron flux and its parametrization function at sea level~\cite{Gord04}.}}
\label{fig:com}
\end{figure}

The early work of Avignone {\em et al.}~\cite{Avig92} showed that isotopes produced in Ge by fast cosmic-ray neutrons could create background in double beta decay and dark matter experiments.
Both the Heidelberg-Moscow~\cite{HM98} and IGEX~\cite{IGEX02} experiments 
observed $\gamma$-rays from such isotopes (e.g. $^{68}Ge$, $^{60}Co$). Thus,
motivated the need to limit the cosmogenic activation, a better understanding of 
the production rate of cosmogenic isotopes is required.
The production rate, $R_i$, of the radioactive 
isotope $i$ can be calculated according to 
\begin{equation}
\label{eq:rate}
R_i = \sum_j N_j \int \phi(E)\sigma_{ij}(E) dE,
\end{equation}
\noindent
where $N_j$ is the number of target nuclear isotope $j$, $\sigma_{ij}$ is the 
neutron excitation function for the product $i$ from target $j$, and $\phi$ 
is the cosmic neutron flux. We ignore the contribution of cosmic protons 
because the flux is much smaller. 
Ziegler carried out a comprehensive study on neutron cosmic ray
flux~\cite{Zieg98} and pointed out that some of the
data from early measurements~\cite{Hess59, Heid71} is incorrect
or of marginal quality~\cite{Pres74}. Several cosmic neutron fluxes~\cite{Hess59, Lalp67} were used in the evaluation of cosmogenic production in both natural 
germanium and enriched $^{76}$Ge conducted by Avignone {\it et al.}~\cite{Avig92} while significant differences emerged 
in the production rate of $^{68}$Ge and $^{60}$Co due to the varying values for the cosmic neutron flux. 
The quality of the neutron flux data has significantly
improved due to new measurements~\cite{Ryan96}. 
Improved recent measurements by Gordon {\it et al.}~\cite{Gord04} show that 
the flux density spectrum at sea level can be parametrized as
\begin{equation}
\label{eq:flux}
\phi(E) = 1.006\times10^{-6}e^{-0.35\ln^2E+2.1451\ln E}+1.011\times10^{-3}e^{-0.4106\ln^2E-0.667\ln E}
\end{equation}
\noindent
where $E$ is neutron kinetic energy in MeV and $\phi$ in units 
of cm$^{-2}$s$^{-1}$MeV$^{-1}$. This parametrization function agrees with the data within $\sim$2\% accuracy as
shown in Fig.~\ref{fig:com}. 

Note that the variation of the cosmic ray muon and neutron fluxes at various locations around world have been 
reported by Ziegler~\cite{Zieg98} and Gordon {\it et al.}~\cite{Gord04}. Cosmic ray muon and neutron fluxes depend
strongly on the altitude and the variation is described as a function of altitude in Ziegler's paper~\cite{Zieg98}.
 In addition, the variation from different locations at the sea level caused by geomagnetic rigidity 
 is also substantial~\cite{Zieg98, Gord04}. In the northern hemisphere, this variation is within 
 10\%~\cite{Gord04}. This work applies the measured cosmic ray neutron flux in the northern 
hemisphere described in Eq.(\ref{eq:flux}). 

\begin{figure}[htb!!!]
\includegraphics[angle=0,width=10.cm] {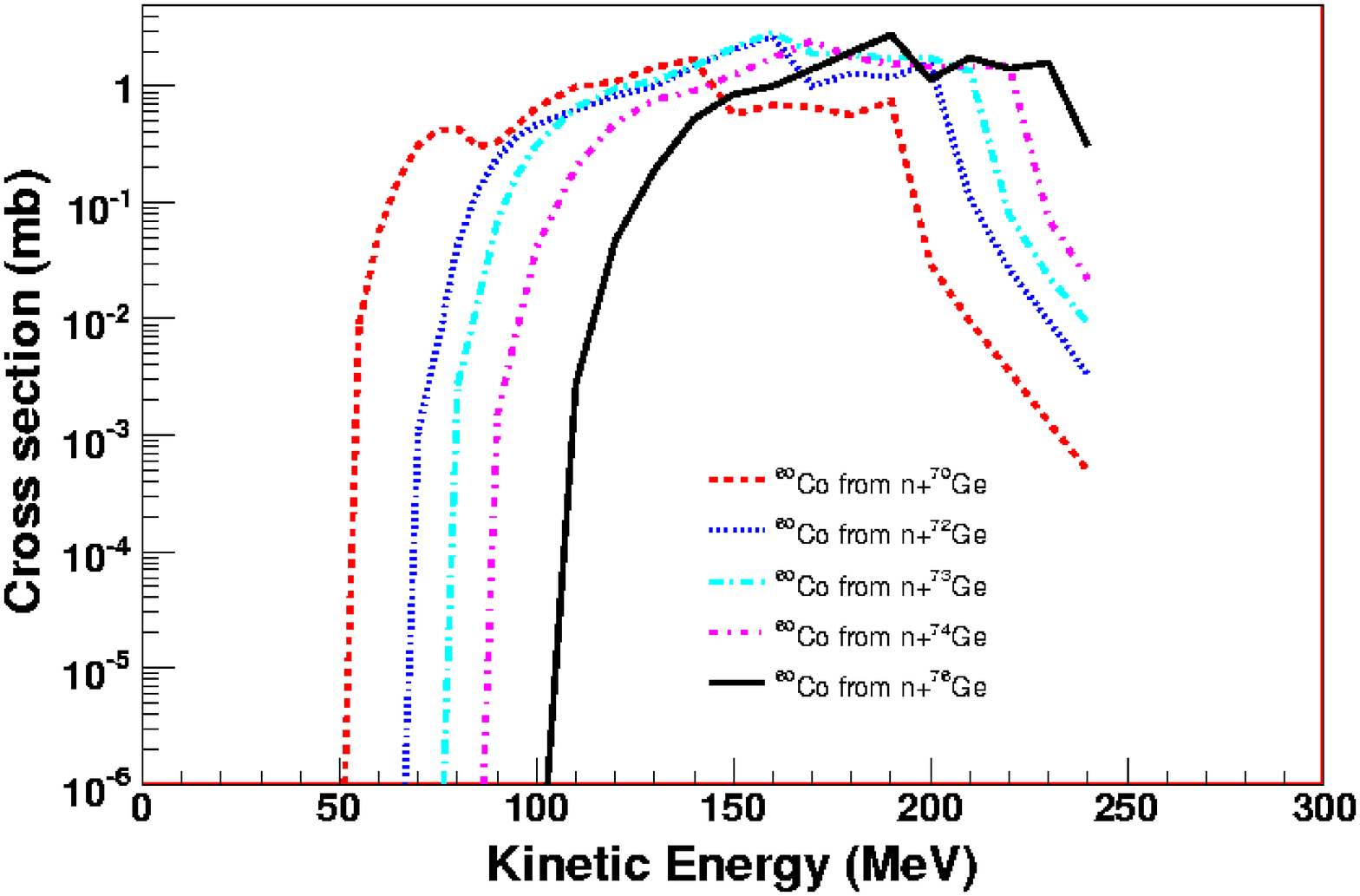}
\caption{\small{$^{60}Co$ production cross section as a function of neutron
kinetic energy.}}
\label{fig:co60}
\end{figure}

%\begin{figure}[htb!!!]
%\includegraphics[angle=0,width=10.cm] {plotGe68.eps}
%\caption{\small{$^{68}Ge$ production cross section as a function of neutron
%kinetic energy.}}
%\label{fig:ge68}
%\end{figure}

\begin{figure}[htb!!!]
\includegraphics[angle=0,width=10.cm] {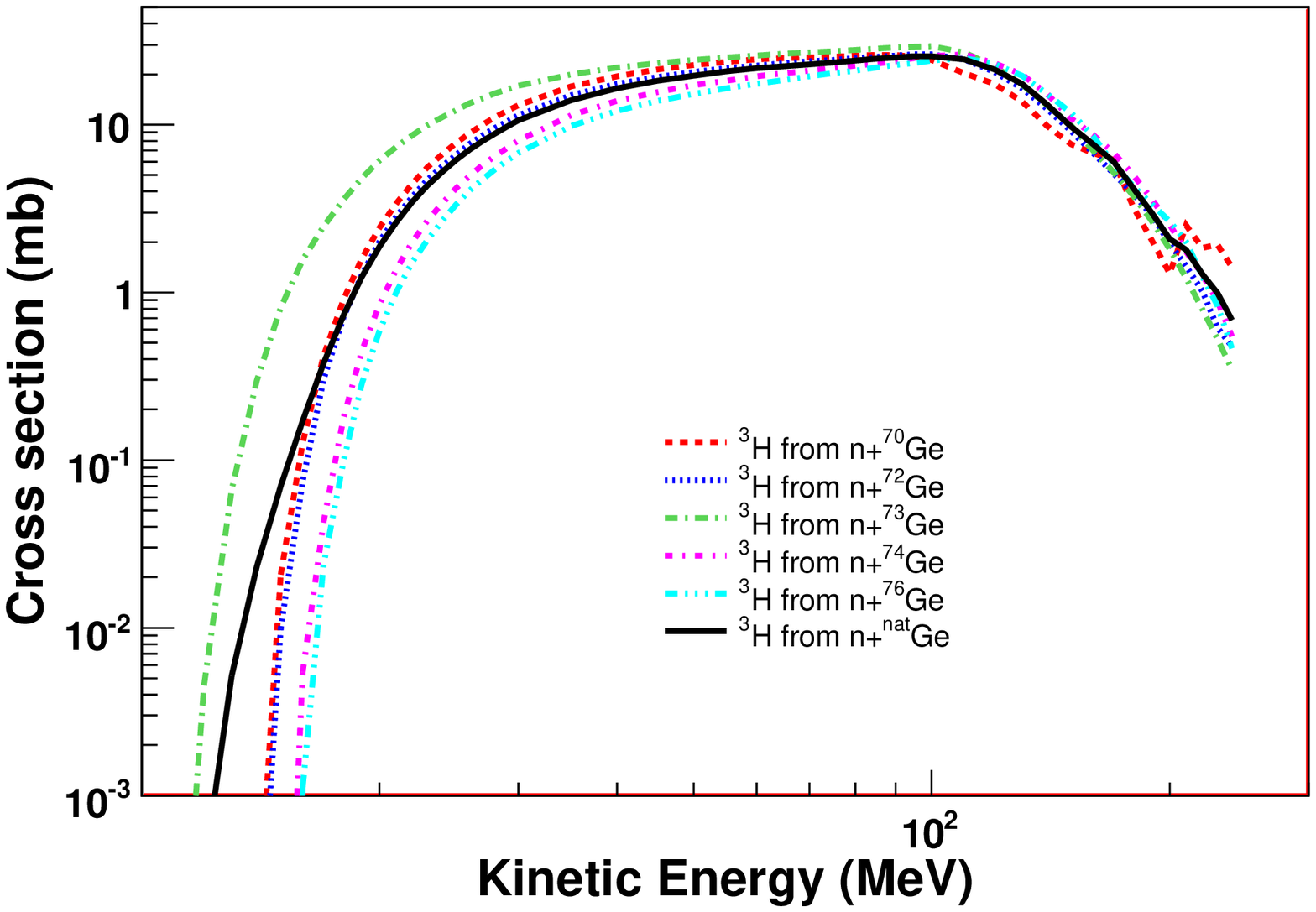}
\caption{\small{$^{3}H$ production cross section as a function of neutron
kinetic energy.}}
\label{fig:H3}
\end{figure}

We use the TALYS code 1.0~\cite{TALYS} to 
generate the excitation functions of isotopes produced 
by neutrons on stable isotopes of germanium
and copper. TALYS is a nuclear reaction program which simulates nuclear reactions 
that involve neutrons, photons, protons, deuterons, tritons, $^{3}$He- and alpha-particles, for target
nuclides of mass 12 and heavier. In the TALYS code 1.0, a suite of nuclear-reaction models has been
implemented into a single code system which enables to evaluate nuclear reactions from the
unresolved-resonance region up to intermediate energies. 
We show, for example, the excitation
functions of $^{60}Co$ 
and $^{3}H$ by neutrons on germanium isotopes 
in Fig.~\ref{fig:co60} and~\ref{fig:H3}, respectively. 
Based on Eq. (\ref{eq:rate}),
the production rates of cosmogenic isotopes in natural 
and enriched (nominally 86\% $^{76}Ge$ and 14\% $^{74}Ge$)
germanium are estimated and tabulated in Table~\ref{tab:rate}. 
Isotopes with half-lives above 10,000 years are not
listed because their contribution to the backgrounds are expected to 
be negligible due to their low decay rate. The production rates of $^{60}Co$ and $^{68}Ge$ in germanium 
agree within a factor of 2 with Ref.~\cite{Bara06, Brod90}, while the rate of $^{3}H$
is a factor of 7 smaller than the estimates in 
Ref.~\cite{Avig92, Klap01},
where the old cosmic neutron flux from Ref.~\cite{Hess59, Lalp67} 
or~\cite{Zieg81} is used. Compared to the old neutron-flux measurements, 
the recently measured flux is smaller
at energies below 50 MeV, but larger between 50 MeV and 1 GeV. 
$^{3}H$ can produce background for dark matter detection via $\beta$-decay
with an end point at 18.6 keV. 
%We illustrate in
%Fig.~\ref{fig:h3decay} the energy distribution of $^{3}H$ 
%decays for one year exposure in a 60 kg enriched germanium detector.
%Although it is possible that the number of $^{3}H$ can be 
%reduced by baking the germanium detector, we will still 
%consider $^{3}H$ as a background source in the following 
%discussion for germanium-based dark matter detection. 

%\begin{figure}[htb!!!]
%\includegraphics[angle=0,width=10.cm] {bkgRate3HPerKgDayKeV.eps}
%\caption{\small{The resulting energy spectrum of cosmogenic $^{3}H$ 
%decays in a 60 kg germanium detector in one year data-taking.
%We assume that the production and transportation
%of the germanium crystals take about 60 days.}}
%\label{fig:h3decay}
%\end{figure}
 
\begin{table}[htb]
\caption{The calculated production rates (per day per kg) of cosmogenic 
isotopes in natural and enriched (86\% $^{76}Ge$ and 14\% $^{74}Ge$) 
germanium using TAYLS and the neutron flux model described in the text. Very long and short lived isotopes are not 
listed because their contribution to the backgrounds are expected to 
be negligible. Also shown are isotopes which are produced by cosmogenic activation
of copper.}
\begin{tabular}{c|r|r|r|r}
\hline
\hline
&\multicolumn{3}{c|}{Production Rate (/(kg day))}& \\
\cline{2-4}
\raisebox{1.5ex}{Cosmogenic Isotope} & Natural $Ge$ & Enriched $Ge$ & Natural $Cu$ &\raisebox{1.5ex}{$t_{1/2}$} \\
\hline
$^{68}Ge$ & 41.3 & 7.2 & & 270.8 d\\
$^{60}Co$ & 2.0 & 1.6 & 46.4 & 5.2714 y\\
$^{57}Co$ & 13.5 & 6.7 & 56.2 & 271.79 d\\
$^{55}Fe$ & 8.6 & 3.4 & 30.7 & 2.73 y\\
$^{54}Mn$ & 2.7 & 0.87 & 16.2 & 312.3 d\\
$^{65}Zn$ & 37.1 & 20.0 & & 244.26 d\\
$^{63}Ni$ & 1.9 & 1.8 &  & 100.1 y \\
$^{3}H$ & 27.7 & 24.0 &  & 12.33 y \\
\hline
\hline
\end{tabular}
\label{tab:rate}
\end{table}

\subsection{An Upper Limit on the Tritium Production Rate in Enriched Ge}
\begin{figure}[htb!!!]
\includegraphics[angle=0, width=10.cm]{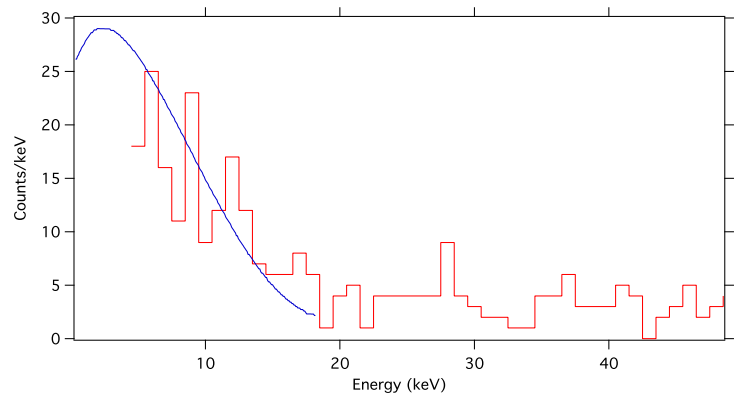}\\
\caption{The low-energy IGEX spectrum (80 kg-d)~\cite{Mor02} with the fit-by-eye to a tritium spectrum.
 \label{fig:spec}}
\end{figure}

For the low-energy data analysis of Ge-detector double-beta decay experiments, the question of the tritium content is
an important issue. The Heidelberg-Moscow~\cite{Bau98} and IGEX~\cite{Mor02} experiments
have both published low-background, low-energy spectra from Ge detectors operated deep underground.
The Heidelberg-Moscow data has a low-energy threshold of 9 keV, whereas the IGEX data
reaches lower to 4 keV. Figure~\ref{fig:spec} shows the IGEX spectrum. Overlaid on that spectrum is the
tritium $\beta$-decay spectrum normalized to 250 counts with a constant background level of 2.5 counts/keV added.
The data is 80 kg-d of exposure. The $"fit"$ was done by eye: that is the normalization of the tritium curve
was determined by adjusting it until it passed approximately through the measured data points.

If one assumes that the normalized curve accurately represents the data, then
the assigned 250 counts represents the maximum number of tritium events that can be contained within 
the data. Since it is likely that other sources of a signal (noise, low energy x-rays, etc.) are present,
we consider the 250 counts to be an upper limit on the spectral contribution due to tritium.

Tritium will be produced within the Ge by high-energy neutron-induced interactions while it is stored above ground.
The Ge detector was underground for about 1 year prior to this data being obtained and the enriched
Ge was above ground 3-5 years after enrichment and prior to going underground~\cite{Avi08}. Therefore
the tritium was produced for 3-5 years and then decayed away for a year before counting.

To convert the 250 counts to an upper limit on the tritium production we use 3 years for the exposure time  as this gives the
largest production rate. The tritium production rate ($k$) depends on the number of counts ($dN$) observed during the counting time ($dt$),  the 
exposure time to cosmic rays ($t_{exp}$)  and the cool-off time before counting starts ($t_{dec}$). The production rate is then given by: 

\begin{eqnarray}
k & = & \frac{dN}{dt} \frac{1}{(1-\exp{\frac{-t_{exp}}{17.79 y}})\exp{ \frac{-t_{dec}}{17.79 y} })} \\
k & < & \frac{250}{80 \mbox{ kg-d}} \frac{1}{(1-\exp{\frac{-3 y}{17.79 y}})\exp{ \frac{-1 y}{17.79 y} })} \\
%k & < & 3.12 \frac{1}{(0.155)(0.945)} \\
k & < & 21 \mbox{ tritium atoms/kg-d}
\end{eqnarray}

\noindent where the mean
life of tritium is 17.79 y.
 This resulting upper limit on the tritium production rate in Ge due to cosmic rays (21 tritium atoms/kg-d) is
 consistent with our TALYS code calculation of 24 tritium atoms/kg-d.
The rate is much lower than a previous calculation~\cite{Avig92} of 110-140 tritium atoms/kg-d, which couldn't discern the fraction of events of $^{72}$Ge(n,x)$^{70}$Ge that led to tritium production.

\section{Comparison between this work and the previous measured rates}
\label{sec:sens}
Avignone {\it et al.} have measured and calculated cosmogenic production rates for some isotopes utilizing 
natural germanium detectors~\cite{Avig92}. Their results are compared to this work in Table~\ref{tab:comparison}.
A reasonable agreement can be found for several isotopes except for $^3{H}$. The differences in the rates of $^3H$ production
is due to the difference in the cosmic neutron energy spectra applied in two calculations. 
\begin{table}[htb]
\caption{The production rate of isotopes in natural germanium.}
\begin{tabular}{c|c|r|c|c}
\hline
\hline
Cosmogenic  & Measured Rate & Calculated Rate &Calculated Rate &Calculate Rate\\
Isotope                &               & (Hess Model~\cite{Hess59})&(Lal Model~\cite{Lalp67})& (This work) \\
                   &(/(kg day))~\cite{Avig92}&(/(kg day))~\cite{Avig92} &(/(kg day))~\cite{Avig92}&(/(kg day)) \\
\hline
$^{3}H$   &-&$\sim$210&$\sim$178&27.7\\
$^{54}Mn$ & 3.3$\pm$0.8& 2.7&0.93& 2.7 \\
$^{65}Zn$ & 38.0$\pm$6.0& 34.4&24.6& 37.1\\
$^{68}Ge$ & 30$\pm$7 &29.6 &22.9&41.3\\
\hline
\hline
\end{tabular}
\label{tab:comparison}
\end{table}

It is worthwhile to mention that our results agree within a factor of two with a recent calculation by Barabanov {\it et al.}~\cite{ibar}. Table~\ref{tab:com} shows the comparison. The difference between two calculations are mainly due to the use of different cosmic ray flux values. 
\begin{table}[htb]
\caption{The production rate of isotopes in germanium}
\begin{tabular}{c|c|r|r|r}
\hline
\hline
Cosmogenic Isotope & \multicolumn{2}{c|} {Natural Ge (/(kg day))}  &\multicolumn{2}{c|} {Enriched Ge (/(kg day))}\\
\hline
                  &Ref.~\cite{ibar}&This work&Ref.~\cite{ibar}&This work\\
\hline
$^{60}Co$ & 2.86 & 2.0 &3.31 &1.6\\
$^{68}Ge$ & 82.7 & 41.3&4.32 &7.2\\
\hline
\hline
\end{tabular}
\label{tab:com}
\end{table}

\section{Cosmogenic production in natural Xe and other targets}
\label{sec:xe}

Liquid noble gases, such as liquid xenon (LXe)~\cite{Xenon10, Bern98, Bern00}
and liquid argon (LAr)~\cite{Boul06, Bene07, Brun05},
have shown excellent pulse shape discrimination capabilities.
Liquid cryogens offer the possibility to construct ton-scale target mass
detectors~\cite{LUX07, ArDM, CLEAN} at a reasonable cost.
We list in Table~\ref{tab:xe} the cosmogenic production rates of isotopes
which can produce potential backgrounds for xenon-based dark matter detection
experiments. Because the interaction rate of WIMPs in xenon decreases
dramatically with the nuclear recoil energy, the sensitive energy
region for xenon-based dark matter experiments is the very low energy
region. Thus, $^{3}H$ $\beta$-decay with the end-point energy of 18.6 keV can
result in large backgrounds for xenon-based dark matter experiments. 
The demonstrated background discrimination power of 1000~\cite{Xenon10} 
via pulse shape analysis is not sufficient to discriminate 
against electronic recoil events induced by $^{3}H$ $\beta$-decay.
However, it is likely that tritium will be reduced by a large factor during purification of the xenon. 

\begin{table}[htb]
\caption{The production rate (per day per kg) of some long lived 
cosmogenic isotopes in natural xenon. Also shown is the half-life. 
Isotopes with very long half-lives or very small production rates are not
listed because their contributions to the background are expected to
be negligible.}
\begin{tabular}{c|c|r}
\hline
\hline
Cosmogenic Isotope & Production Rate (/(kg day)) & $t_{1/2}$\\
\hline
$^{3}H$ & 16.0 & 12.33 y \\
$^{121m1}Te$ & 11.7 & 154 d\\
$^{123m1}Te$ & 12.1 & 119.7 d\\
$^{127m1}Te$ & 5.0 & 109 d \\
$^{101}Rh$ & 0.04 & 3.3 y\\
$^{125}Sb$ & 0.04 & 2.7582 y\\
$^{119m1}Sn$ & 0.02 & 293.1 d\\
$^{123}Sn$ & 0.004 & 129.2 d\\
$^{109}Cd$ & 3.2 & 462.6 d\\
$^{113m1}Cd$ & 0.002 & 14.1 y\\
\hline
\hline
\end{tabular}
\label{tab:xe}
\end{table}

In Table~\ref{tab:h3} we tabulate the cosmogenic 
production rate of $^{3}H$ in various
targets of dark matter detection experiments. These numbers can be used
to guide the requirements for electron-recoil rejection and cryogenic purification to prevent $^3$H
background.

\begin{table}[htb]
\caption{The cosmogenic production rate (per day per kg) of $^{3}H$ 
in various targets.}
\begin{tabular}{c|cccccc}
\hline
\hline
Target &Ar & $Xe$ & $NaI$ & $CsI$ & $TeO_2$ & $CaWO_4$\\
\hline
Rate (/kg/day)&44.4 & 16.0 & 31.1 & 19.7 & 43.7 & 45.5\\
\hline
\hline
\end{tabular}
\label{tab:h3}
\end{table}

It is worthwhile to mention that it has been demonstrated that liquid argon
can achieve a pulse shape discrimination power of $1.3\times 10^{6}$ 
against electronic recoil events at energy of 15 keV~\cite{Lipp08}. 
Thus, at such a high energy threshold, the cosmogenic $^{3}H$ 
will not be a problem for LAr-based dark matter detection and 
the dominant background source will be $^{39}Ar$ contained in natural 
argon.

\section{Conclusions}
\label{sec:con}

We have investigated the cosmogenic production of various isotopes in several
target or source materials pertinent for dark matter and double-beta decay experiments.
The tritium production in these materials due to cosmic-ray neutrons is substantial
and steps must be taken to either reduce exposure of the target to cosmic rays, reduce the resultant $^3$H within the target after exposure,
or develop an event-by-event analysis to remove $^3$H decay events from the data stream.

\section*{Acknowledgments}
We thank Y. D. Chan, J. A. Detwiler, John Wilkerson,  R. Henning and other {\sc{Majorana}}
collaborators for discussion. We thank F. T. Avignone for a careful reading of this manuscript. This work was supported in 
part by NSF grant PHY-0758120 and by the Office of Research at The University
of South Dakota and by Laboratory Directed Research
and Development at Los Alamos National Laboratory. Z.B. Yin
was also partly supported by MOE of China under project No. IRT0624
and the NFSC under grant No. 10635020.

% \appendix

% \section{}
% \label{}

\end{document}